\documentclass[twocolumn,superscriptaddress, secnumarabic,amssymb, amsmath, nobibnotes, aps, prd]{revtex4-2}
\usepackage[hidelinks]{hyperref}
\usepackage{xcolor}
\hypersetup{
    colorlinks,
    linkcolor={black},
    citecolor={blue!50!black},
    urlcolor={blue!80!black}
}
\usepackage{graphicx}
\usepackage{float}
\usepackage{mathtools}
\usepackage[flushleft]{threeparttable}
\usepackage{subcaption}
\usepackage{orcidlink}

\usepackage{color}
\usepackage{soul}

\usepackage[normalem]{ulem}

\def\prd{PRD }

\begin{document}

\title[Interaction between linear polarized plane gravitational waves and a plane electromagnetic wave]{Interaction between linear polarized plane gravitational waves\\ and a plane electromagnetic wave in the electromagnetic-gravity analogue}

\author{Enderson Falcón-Gómez\orcidlink{0000-0002-0008-0624}}%
\affiliation{Signal Theory and Communication Department. University Carlos III of Madrid. Leganés, Spain}
\author{Vittorio De Falco\orcidlink{0000-0002-4728-1650}}%
\affiliation{Scuola Superiore Meridionale,  Largo San Marcellino 10, 80138 Napoli, Italy}
\affiliation{Istituto Nazionale di Fisica Nucleare, Sezione di Napoli, Via Cintia 80126 Napoli, Italy}
\author{Kerlos Atia Abdalmalak\orcidlink{0000-0002-9544-2412}}
\affiliation{Signal Theory and Communication Department. University Carlos III of Madrid. Leganés, Spain}
\affiliation{Electrical Engineering Department, Aswan University, Aswan 81542, Egypt}
\author{Adrián Amor-Martín\orcidlink{0000-0002-6123-4324}}
\affiliation{Signal Theory and Communication Department. University Carlos III of Madrid. Leganés, Spain}
\author{Valentín De La Rubia\orcidlink{0000-0002-2894-6813}}
\affiliation{Departamento de Matemática Aplicada a las TIC, ETSI de Telecomunicación, Universidad Politécnica de Madrid, Madrid, Spain}
\author{Gabriel Santamaría-Botello\orcidlink{0000-0003-4736-0030}}%
\affiliation{Department of
 Electrical, Computer, and Energy Engineering, University of Colorado Boulder, Boulder, CO, USA}
\author{Luis Enrique García Muñoz\orcidlink{0000-0002-3619-7859}}%
\email{legarcia@ing.uc3m.es}
\affiliation{Signal Theory and Communication Department. University Carlos III of Madrid. Leganés, Spain}

\begin{abstract}
We study the interaction among gravitational and electromagnetic plane waves by means of an analogue electromagnetic model of gravity, where the gravitational properties are encoded in the electromagnetic properties of a material in flat space-time. In this setup, the variations in the metric tensor produced by the gravitational waves are codified as space-time-varying electromagnetic properties. We used an in-housed code based in the finite-difference time domain method to conduct numerical experiments, where we found that, when a monochromatic gravitational plane wave interacts with a narrow-band electromagnetic plane wave, an infinite number of sidebands, equally separated between themselves, are induced by the gravitational wave. 
Finally, we discuss possible future applications of this effect as an alternative method to directly detect gravitational waves.
\end{abstract}

\date{\today}%

\maketitle

\section{Introduction}
\label{sec:intro}
Gravitational waves (GWs) have been directly observed for the first time in September 2015, validating thus the predictions of General Relativity (GR) \cite{Abbott2016,Abbott2017GW170817}. Since then, the interest in their measurement has considerably increased, because they are ideal footprint for compact binary mergers \cite{Ciolfi2020,Pian2021,Wysocki2022} and can provide fundamental information on the gravitational interaction \cite{Buonanno2014aza,Schmidt2020}. Furthermore, astrophysical GWs are normally emitted very far from the Earth, and on this long travel they do not undergo, in good approximation, either scattering or modifications in their properties \cite{bailes2021gravitational}.

Despite that, during the last decades, the sensitivity of the current interferometers, such as Advanced LIGO \cite{aasi2015advanced}, Advanced Virgo \cite{acernese2014advanced}, and KAGRA \cite{aso2013interferometer}, has been strongly improved, it is always worth exploring alternative techniques for GW detection, that could foster new observational schemes. A promising approach is based on the developments of analogue models of gravity \cite{Barcelo2011}. They represent original strategies to model various astrophysical phenomena in GR via in-hand physical systems, which allows us to provide feasible laboratory experiments. Among the different proposals, it is worth mentioning: Zel'Dovich and collaborators studied a GPW with frequency $f_{\rm GW}$ inside a crystal slab that excites an acoustic wave, which in turn, produces scattering of a photon of frequency $f_{\rm e}$ and generates thus two sidebands $f_e \pm f_{\rm GW}$ \cite{kopvillem1973detection}; Brodin carried out a study on the interaction of a GW with a magnetized multicomponent plasma medium, obtaining that the electromagnetic wave (EMW) is modulated by density gradients created inside the plasma due to the influence of GWs \cite{brodin2001photon}; Li considered the direct interaction of light propagating in vacuum as a Gaussian beam and discovered two modulated sidebands in the presence of GWs \cite{li2003electromagnetic}, similar to Zel'Dovich's work; Servin analysed the resonant interaction between GWs and EMWs in a plasma medium \cite{servin2003resonant}; Gertsenshtein estimated the extent of GW excitation by light \cite{gertsenshtein1962wave}; Jones and Singleton focused on GW propagating in vacuum and the consequent production of electromagnetic (EM) radiation arising from it \cite{jones2019interaction}, which is small but not null. However, higher-order sidebands in the EMW's spectrum are usually neglected in the bibliography and their behavior needs to be further inquired to understand their relationship with GWs.
{EM-based detection systems are a promising tool to develop new detection methods
to search for new and more GW events. In this context, resonant cavities (structures that store and enhance EM-field magnitude) have been proposed as medium for detecting GWs, which is based on the study of the interaction between gravitational and EM fields \cite{Mik2019,Berlin2022}. This method has been analysed and upgraded in several studies \cite{pegoraro1978electromagnetic,pegoraro1978operation,Berlin2022}. These progresses are fundamental not only for astrophysical purposes \cite{Berlin2022}, but also for cosmological reasons \cite{Herman2022}. }

In this manuscript, we study the effects that a monochromatic gravitational plane wave (GPW) of frequency $f_{\rm GW}$ produces on a narrow-band electromagnetic plane wave (EPW) with carrier frequency $f_{\rm e}$. To accomplish this, we use an analogue model for gravity, where we encode the gravitational field into the EM properties of a material in flat space-time \cite{plebanski1960electromagnetic}. In this framework, the problem turns into the interaction of an EM wave with a material endowed with varying space-time EM properties \cite{taravati2018giant}, that can be solved numerically via the Finite Differences Time-Domain (FDTD) method \cite{taflove2005computational}. We show that the interaction of a GPW with an EPW triggers the multiple scattering of photons by gravitons, modifying thus the spectrum of the EMWs by generating an infinite number of sidebands with frequencies $f_{\rm n} = f_e + n f_{\rm GW}$, being $n$ an integer number. This effect is quite similar to the Brillouin cascade scattering effect occurring in photonic crystal fibers \cite{wolff2017cascaded}. This article is organized as follows: in Sec. \ref{section:maxwell_equations_in_curved_spacetime}, we introduce the analogue EM-model of gravity, as well as our notations; in Sec. \ref{sec:methods}, the simulation methodology is devised; in Sec. \ref{section:results_and_discussion}, we present and discuss the results underlying the effects of GPWs on the EM spectrum considering different EPW polarizations; finally in Sec.\ref{section:conclusion}, we draw the conclusions and outline future perspectives.

\emph{Notation.} We indicate with $g_{\mu\nu}$ a symmetric metric tensor with signature $(-,+,+,+)$, inverse metric $g^{\mu\nu}$, and determinant $g=\det(g_{\mu \nu})$. Greek indices run as $0,1,2,3$, and Latin indices run as $1,2,3$. We denote vectors as $\vec{A}$ and tensors as $\overleftrightarrow A$, defined component-wise as $A_{\mu \nu}$. Scalar product, vector product, and tensor-vector multiplication are indicated with $\vec{A} \cdot \vec{B}$, $\vec{A}\times \vec{B}$, and $\overleftrightarrow{A}\vec{B}$, respectively. The differential calculus is performed in a flat space-time throughout all the paper. The vector basis adapted to the spatial cartesian coordinate system is given by the orthonormal vectors: $\hat{e}_{\rm x},$ $\hat{e}_{\rm y},$ and $\hat{e}_{\rm z}$. $\epsilon _0$, $\mu_0$, and $c$ are the dielectric permittivity, magnetic permeability, and speed of light in the vacuum, respectively. In the EM field notations, calligraphic symbols, e.g., $\vec{\mathcal{E}}$ and $\vec{\mathcal{B}}$, refer to quantities in time domain, while normal ones, e.g., $\vec{E}$ and $\vec{B}$, are set in the frequency domain.

\section{Methods}
\label{sec:methods}
In this section, we introduce the fundamental concepts and methods exploited in our work. First of all, in Sec.~\ref{section:maxwell_equations_in_curved_spacetime}, we present the Maxwell equations including the effect of a generic curved space-time within the EM-gravity analogue model. In Sec.~\ref{sec:measurements_in_curved_spacetime}, the tetrad-formalism is taken into consideration for performing the measurements of physical observables in curved space-times, and finally, in Sec.~\ref{sec:fdtdalgorithm}, we describe the FDTD method. To conclude, in Sec.~\ref{sec:simulation_conditions}, we explain the experimental setup and procedures employed for carrying out the numerical simulations.

\subsection{Maxwell equations}
\label{section:maxwell_equations_in_curved_spacetime} 
The propagation of the EM radiation in a gravitational field is influenced by the geometric curved background. This interaction was studied by {Plebanski} in 1960 \cite{plebanski1960electromagnetic}, who introduced an analogue EM-model of gravity to describe EM wave propagation in gravitational fields, governed by the Maxwell equations
\begin{subequations}
\begin{align}
&\nabla \cdot \vec{\mathcal{D}}(\vec{r},t)=\sqrt{-g}\ \mathcal{\sigma}(\vec{r},t),\label{eq:subsection:maxwell_equatinos_in_non-covariant_notation17}\\
&\nabla \cdot \vec{\mathcal{B}}(\vec{r},t) = 0,
\label{eq:subsection:maxwell_equatinos_in_non-covariant_notation04}\\
&\nabla \times \vec{\mathcal{E}}(\vec{r},t) = -\frac{\partial \vec{\mathcal{B}}(\vec{r},t)}{\partial t},\label{eq:subsection:maxwell_equatinos_in_non-covariant_notation08}\\
&\nabla \times \vec{\mathcal{H}}(\vec{r},t)=\sqrt{-g}\  \vec{\mathcal{J}}(\vec{r},t)+\frac{\partial  \vec{\mathcal{D}}(\vec{r},t) }{\partial t},
\label{eq:subsection:maxwell_equatinos_in_non-covariant_notation16}
\end{align}
\label{eq:subsection:maxwell_equatinos_in_non-covariant_whole}
\end{subequations}
together with the related constitutive equations
\begin{subequations}
\begin{align}
\vec{\mathcal{D}}(\vec{r},t) &= {\overleftrightarrow \epsilon}(\vec{r},t) \vec{\mathcal{E}}(\vec{r},t)+\vec{\Gamma}(\vec{r},t) \times \vec{\mathcal{H}}(\vec{r},t),\label{eq:subsection:constitutive_relationship01}\\
\vec{\mathcal{B}}(\vec{r},t) &= {\overleftrightarrow \mu}(\vec{r},t) \vec{\mathcal{H}}(\vec{r},t)+ \vec{\mathcal{E}}(\vec{r},t) \times \vec{\Gamma}(\vec{r},t),\label{eq:subsection:constitutive_relationship02}
\end{align}
\label{eq:subsection:constitutive_relationships}
\end{subequations}
being $\mathcal{\sigma}$ the charge density, $\vec{\mathcal{D}}$ the electric flux density, $\vec{\mathcal{B}}$ the magnetic flux density, $\vec{\mathcal{\mathcal{E}}}$ the electric field intensity, $\vec{\mathcal{H}}$ the magnetic field intensity, $\vec{\mathcal{J}}$ the current density, $\overleftrightarrow \epsilon$ the dielectric permittivity tensor, $\overleftrightarrow \mu$ the magnetic permeability tensor, and finally,  $\vec{\Gamma}$ a magnetoelectric coupling term. Furthermore, the tensors $\overleftrightarrow \epsilon$ and $\overleftrightarrow \mu$ together with the vector $\vec{\Gamma}$ are related to the metric tensor $g_{\mu \nu}$ via 
\begin{subequations}
\begin{align}
\frac{\left({\overleftrightarrow \epsilon}\right)^{ij}}{\epsilon_0} = \frac{\left({\overleftrightarrow \mu}\right)^{ij}}{\mu_0} &= -\frac{\sqrt{-g}}{g_{00}}g^{ij},\label{eq:subsection:Constitutive_relationships_for_non-covariant_Maxwell_equations_in_curved_spacetime_C}\\
{\vec{\Gamma} }_k &= \frac{1}{c}\frac{g_{0k}}{g_{00}}.
\label{eq:subsection:Constitutive_relationships_for_non-covariant_Maxwell_equations_in_curved_spacetime_D}
\end{align}
\end{subequations}

It is important to note that Eqs. \eqref{eq:subsection:maxwell_equatinos_in_non-covariant_whole} and \eqref{eq:subsection:constitutive_relationships} are framed in a flat space-time. The flat space-time perturbed by a GPW is modeled via the space-time varying tensors $\overleftrightarrow \epsilon$ and $\overleftrightarrow \mu$. The GWs are framed within the linearized GR theory and mathematically can be written as $g_{\mu \nu} = \eta_{\mu \nu}+h_{\mu \nu}$, where $h_{\mu \nu}$ represents the perturbation tensor on the flat Minkowski space-time, $\eta_{\mu \nu}$ \cite{maggiore2008gravitational,landau2013classical,carroll2019spacetime}. 

For a GPW propagating along the $z$-axis (see Eq. (1.34) in \cite{maggiore2008gravitational}), Eq. \eqref{eq:subsection:Constitutive_relationships_for_non-covariant_Maxwell_equations_in_curved_spacetime_C}, after a manipulation, becomes 
\begin{equation}
\frac{\overleftrightarrow \epsilon}{\epsilon_0} = \frac{\overleftrightarrow \mu}{\mu_0}  =  \begin{pmatrix}
1-f_\oplus & -f_\otimes  & 0   \\
-f_\otimes & 1+f_\oplus & 0   \\
0 & 0 & 1  \\
\end{pmatrix},
\label{eq:subsection:Constitutive_relationships_for_non-covariant_Maxwell_equations_in_curved_spacetime_F}
\end{equation}
where 
\begin{subequations}
\begin{align}
f_{\oplus}(z,t) &= h_{\oplus} \cos\left( \omega_{\rm GW} t-\kappa_{\rm GW}z\right),\label{eq:functionplus}\\
f_{\otimes} (z,t) &= h_{\otimes} \cos\left( \omega_{\rm GW} t-\kappa_{\rm GW} z+ \Delta \phi\right),
\label{eq:functiocross}
\end{align}
\end{subequations}
$h_{\oplus}$ and $h_{\otimes}$ being the amplitudes of the polarization states ``plus'' and ``cross'' of the GPW, respectively, $\Delta \phi$ the phase difference between them, $\omega_{\rm GW}$ the GW angular frequency, and $\kappa_{\rm GW} = \omega_{\rm GW}/c$ the wave number. Finally, we stress that the Maxwell equations \eqref{eq:subsection:maxwell_equatinos_in_non-covariant_whole} are expressed in the \emph{standard (spherical) coordinate system}.

\subsection{Tetrad formalism and measurements of observables in curved space-times}
\label{sec:measurements_in_curved_spacetime}
In GR, it is extremely important to assign a precise meaning to the measured observables (e.g., frequency, energy, EM fields) and this can be achieved by resorting to the (strong) Equivalence Principle. This can be successfully accomplished by employing the Local Inertial Frame (LIF) coordinates. Therefore, we will need to translate calculations performed from the standard coordinate system to the LIF coordinates. 

To this purpose, we use the tetrads $\tau _{(\alpha)} ^\mu$ (LIF vector basis), where $\mu$ represents the tetrad-component index, whereas $(\alpha)$ is a tetrad-vector basis index. They must fulfill the following equation (see Ref. \cite{Capozziello2022}, for details)
\begin{equation} \label{eq:tetradprop}
 g_{\mu \nu} \tau _{(\alpha)} ^\mu\tau _{(\beta)} ^\nu = \eta _{\alpha \beta}.
\end{equation}
For an observer at rest in the background of a GPW, the tetrads assume the following form \cite{maggiore2008gravitational}
\begin{subequations}
\begin{align}
\tau _ {(0)}&= \left(1,0,0,0\right),
\label{eq:tetrad0}\\
\tau _ {(1)}&= \left(0,\frac{h+f_\oplus}{D_1},\frac{f_\otimes}{D_1},0\right),\label{eq:tetrad1}\\
\tau _ {(2)}&= \left(0,-\frac{h-f_\oplus}{D_2},\frac{f_\otimes}{D_2},0\right),\label{eq:tetrad2}\\
\tau _ {(3)} &= \left(0,0,0,1\right).\label{eq:tetrad3}
\end{align}
\end{subequations}
 where 
\begin{subequations}\label{eq:denominator_tetra_whole}
\begin{align}
D_1 &= \sqrt{1+h}\sqrt{\left(h+f_\oplus\right)^2+f_\otimes ^2},\label{eq:denominator_tetra_02}\\
D_2 &= \sqrt{1-h}\sqrt{\left(h-f_\oplus\right)^2+f_\otimes ^2},\label{eq:denominator_tetra_03}\\
h &= \sqrt{f_{\oplus}^2+f_{\otimes}^2}.\label{eq:denominator_tetra_04}
\end{align}
\end{subequations}

To project $\vec{\mathcal{E}}$ and $\vec{\mathcal{B}}$ in the LIF, we exploit the tetrads to raise and lower the indices of the EM tensor field, 
\begin{equation}
F^{\mu\nu}= \begin{pmatrix}
0 & -\mathcal{E}_x & -\mathcal{E}_y & -\mathcal{E}_z \\
\mathcal{E}_x & 0 & -\mathcal{B}_z & \mathcal{B}_y\\
\mathcal{E}_y & \mathcal{B}_z & 0 & \mathcal{B}_x\\
\mathcal{E}_z & -\mathcal{B}_y & \mathcal{B}_x & 0
\end{pmatrix},    
\end{equation}
as follows $F _{\alpha \beta}=F_{\mu \nu}\tau _{(\alpha)}^\mu \tau _{(\beta)} ^\nu$ \cite{carroll2019spacetime}.

\subsection{The FDTD method}
\label{sec:fdtdalgorithm}
The analysis of the interaction between the GW and the EM field is performed by employing the FDTD method \cite{taflove2005computational}, which is a full-wave numerical technique that has been already used to deal with the propagation of EM fields in space-time-varying materials (see e.g., \cite{taravati2018giant,Stewart2018finite,scarborough2021efficient,mirmoosa2019time,kord2019magnetless} and references therein). 
\begin{figure}
\centering
\includegraphics[scale=0.3]{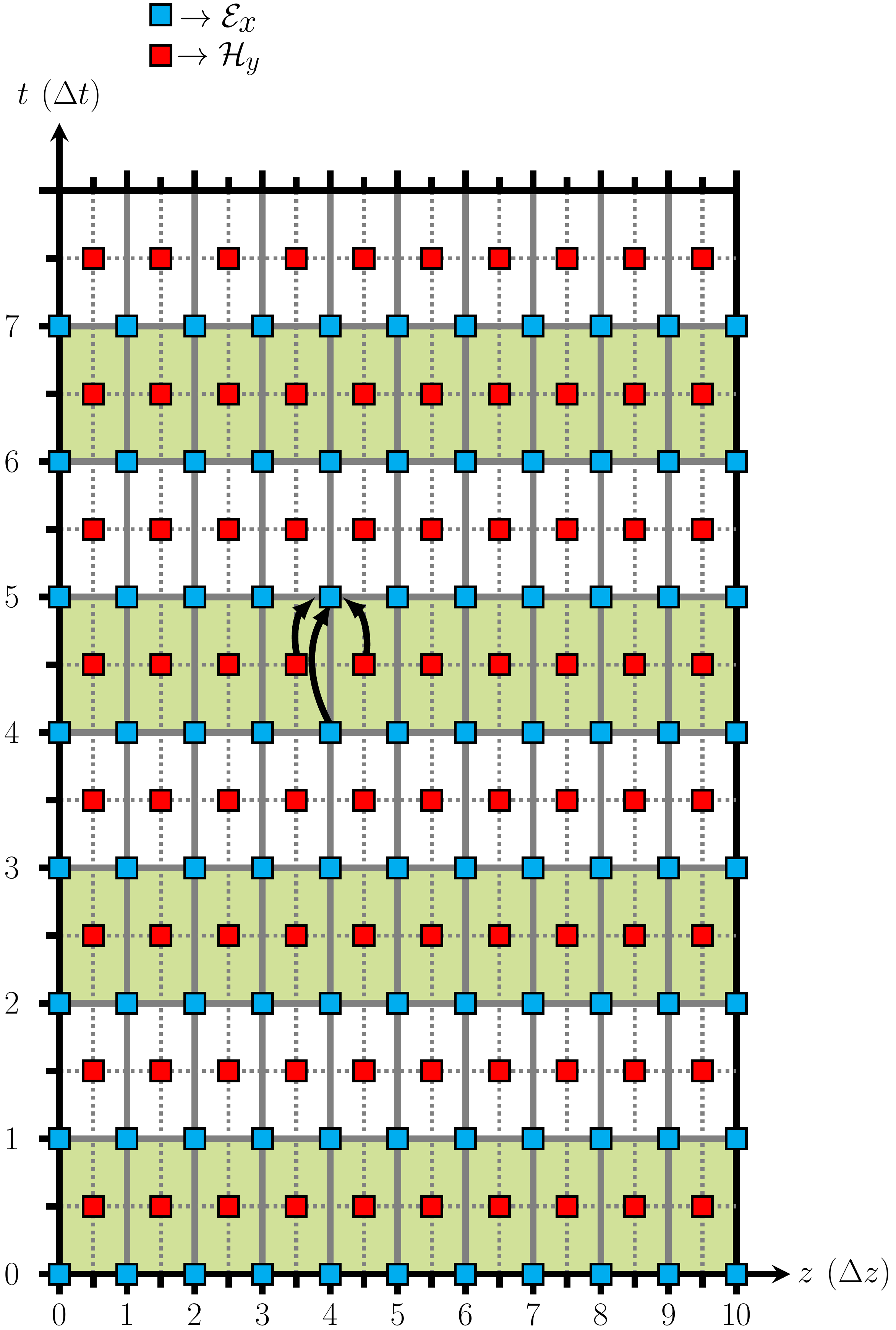}
\caption{Space-time solving scheme diagram of the FDTD algorithm. Green- and white-colored sets represent the computational domains at even and odd time-steps, respectively. Cyan and red squares represent the electric, $\mathcal{E}_x$, and magnetic, $\mathcal{H}_y$, fields, respectively.}
\label{fig:fdtd_algorithm}
\end{figure}

Since we need to describe the GW and EMW propagations in GR, which are both only transversal \cite{maggiore2008gravitational}, our problem configures to be one-dimensional in space, which we assume to be directed along the $z$-axis. This is a fundamental information to prepare the numerical grid for applying the FDTD method. We use an example to explain how our numerical implementation works. To this end, we consider the following one-dimensional problem
\begin{equation}
\frac{\partial \mathcal{E}_x}{ \partial t} = -\frac{1}{\epsilon _0 }\frac{\partial \mathcal{H}_y}{ \partial z}.
\label{eq:plane_wave_equation}
\end{equation}

In the FDTD, both space and time domains are uniformly discretized into cells of length $\Delta z$ and steps of duration $\Delta t$ \cite{taflove2005computational}, respectively. In Fig.~\ref{fig:fdtd_algorithm} we display, as an example, a space-time lattice of a hypothetical FDTD simulation, in which the spatial domain consists of 10 cells along the $z$-axis and the algorithm runs over 7 time-steps (without considering the initial instant $t = 0$). Green- and white-colored rectangles represent the grids at even and odd time steps, respectively.

Electric and magnetic fields are interleaved in space-time as follows: the electric field (cyan squares in Fig.~\ref{fig:fdtd_algorithm}) is evaluated at the edges of the cells and at the beginning of each time step, whereas the magnetic field (red squares in Fig.~\ref{fig:fdtd_algorithm}) is computed at the middle of each cell and in between time steps. We define $z_{k} = k \Delta z$ and $t_n = n \Delta t$ to indicate the position and time steps, respectively, with $n,k$ being positive integers, in which the electric field components are evaluated; whereas the magnetic field components are calculated in the points $z_{k \pm \frac{1}{2}} = \left(k\pm \frac{1}{2} \right) \Delta z$  and $t_{n \pm \frac{1}{2}} = \left( n\pm\frac{1}{2} \right) \Delta t$. 

Following this discretization pattern and interleaving of EM field components in space-time, we approximate the first-order derivatives with respect to both space and time in the Maxwell equations \eqref{eq:subsection:maxwell_equatinos_in_non-covariant_whole} via \emph{central finite differences} \cite{taflove2005computational}. For example, for approximating the terms $\frac{\partial \mathcal{H}_y}{\partial z}$ and $\frac{\partial \mathcal{E}_x}{\partial t}$ in Eq. \eqref{eq:plane_wave_equation} in the generic space-time point $(z_{\rm k},t_{\rm n + \frac{1}{2}})$, we apply
\begin{subequations}
\begin{align}
 \frac{\partial \mathcal{E}_x}{\partial t}\left(z_{\rm k},t_{\rm n + \frac{1}{2}}\right) &= \frac{\mathcal{E}_x \left(z_{\rm k},t_{\rm n+1}\right)
 - \mathcal{E}_x \left(z_{\rm k},t_{\rm n}\right)}{\Delta t},
\label{eq:central_differences_approximation_Exdet}\\
 \frac{\partial \mathcal{H}_y}{\partial z}\left(z_{\rm k},t_{\rm n + \frac{1}{2}}\right) &= \frac{\mathcal{H}_y \left(z_{\rm k + \frac{1}{2}},t_{\rm n + \frac{1}{2}}\right)
 - \mathcal{H}_y \left(z_{\rm k- \frac{1}{2}},t_{\rm n + \frac{1}{2}}\right)}{\Delta z}.
\label{eq:central_differences_approximation_Hydez}
\end{align}
\end{subequations}

Finally, upon substituting Eqs. \eqref{eq:central_differences_approximation_Exdet} and \eqref{eq:central_differences_approximation_Hydez} into Eq. \eqref{eq:plane_wave_equation}, we update $\mathcal{E}_x$ employing electric and magnetic field components calculated at the previous steps, namely
\begin{align}
\mathcal{E}_x\left(z_{k},t_{\rm n+1}\right) &=\mathcal{E}_x\left(z_{k},t_{\rm n}\right)\notag\\ 
&+\frac{\mathcal{H}_y\left(z_{k - \frac{1}{2}},t_{\rm n + \frac{1}{2}}\right)-\mathcal{H}_y\left(z_{k + \frac{1}{2}},t_{\rm n + \frac{1}{2}}\right)}{\epsilon _0\Delta z  / \Delta t}.
\label{eq:example_of_maxwell_equation_discretized}
\end{align}
The strategy to deal with the above equation is better illustrated in Fig.~\ref{fig:fdtd_algorithm} for $\mathcal{E}_x\left(z_{4},t_{\rm 4}\right)$. We note how previous values of EM field components permit to determine their future values (so-called \emph{causal structure}). The EM field components are initialized with null values and the two electric field components at the boundaries of the spatial domain, i.e., $z_{0}$ and $z_{10}$, are never updated, therefore, we say that the computational domain is surrounded by two Perfect Electric Conductor (PEC) flat boundaries (extended on $x-y$ plane) at each end \cite{taflove2005computational}. The tangential component of an electric field, after having assigned a PEC boundary condition, is always null, and this will generate reflections at the two extremes, since a PEC behaves as a perfect reflector.

Furthermore, at this point it is worth mentioning that, in our numerical experiments, we have used a Generalized Material Independent Perfectly Matched Layer (GMIPML) \cite{zhao1998generalized} to avoid the reflections produced at both ends of the computational domain, whereby we could simulate a problem in open space where radiation never gets reflected back and ``propagates'' towards the infinity once it reaches the GMIPML, even though a finite grid is used. The GMIPML is a layer of finite thickness that is placed next to the boundaries of the computational domain and consists of a material with losses that absorbs the incoming radiation similar to the absorbers used in the anechoic chambers in radiation or acoustic environments (see Ref. \cite{zhao1998generalized}, for more details). To introduce energy in the computational domain, we use a \emph{soft source} \cite{taflove2005computational}, which requires the addition of a function $f=f(t)$ to the electric field at a cell $k_s$ in the grid as follows: $\mathcal{E}_x (z_{k_s},t_n) = \mathcal{E}_x (z_{k_s},t_n) + f(t_n)$. A more detailed description of the FDTD algorithm can be found in the Taflove's work \cite{taflove2005computational}. Implementing all the aforementioned numerical procedures we develop our version of the FDTD solver, which has been validated against the analytic time-varying problems reported in Ref. \cite{morgenthaler1958velocity}. 
\begin{figure*}[ht!]
\centering
\includegraphics[scale=0.5]{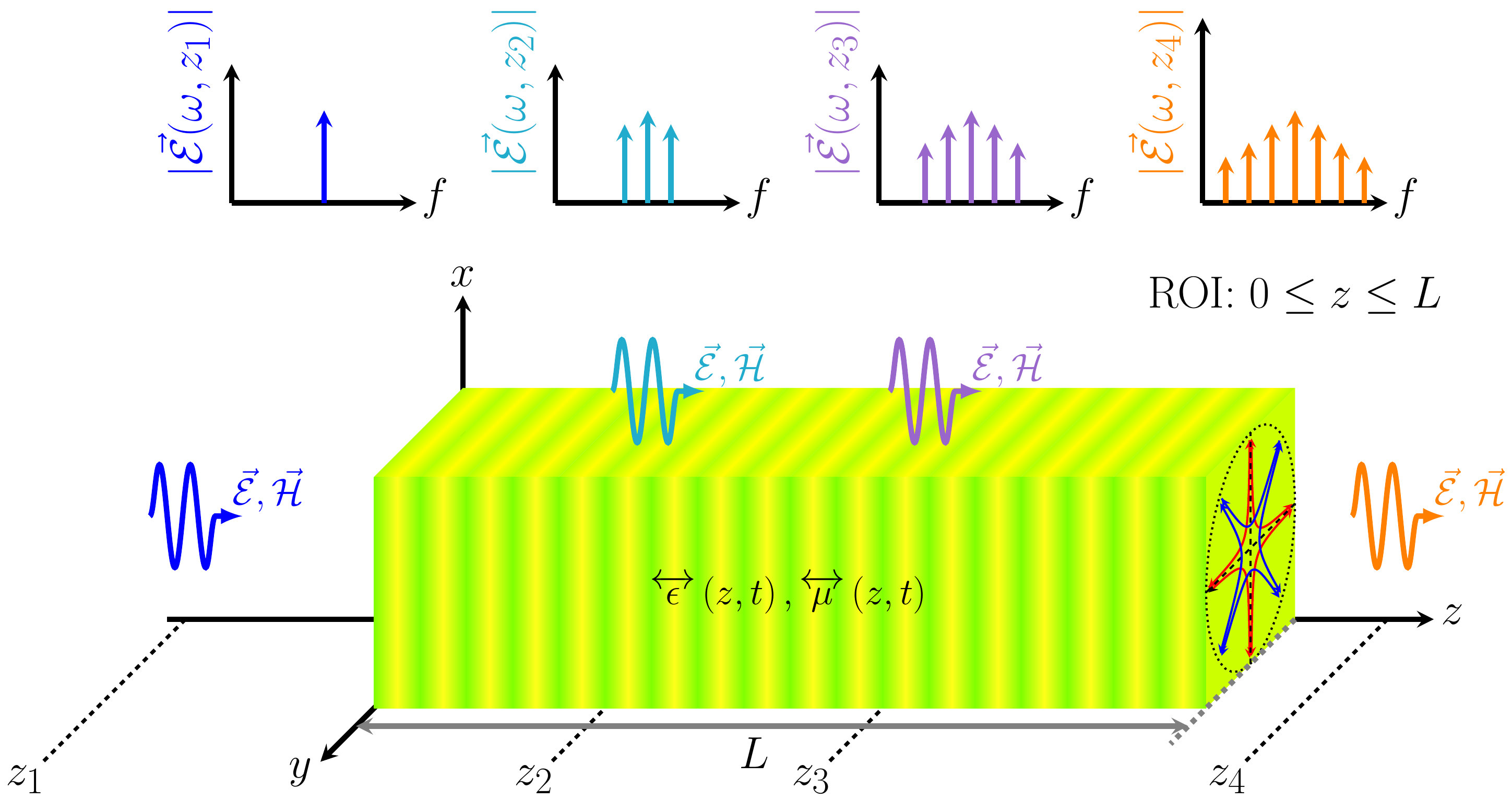}
\caption{Schematic representation of the interaction between a GPW with arbitrary polarization states, and an EPW, both propagating along the $z$-axis. The yellow-green cuboid represents an infinite (metallic) slab (denoted here as ROI), whose EM properties are described by $\overset{\text{\tiny$\leftrightarrow$}}{\epsilon} (z,t)$ and $\overset{\text{\tiny$\leftrightarrow$}}{\mu} (z,t)$ (cf. Eq. \eqref{eq:subsection:Constitutive_relationships_for_non-covariant_Maxwell_equations_in_curved_spacetime_F}) and thickness $L$. It is placed with its section on top of the $x-y$ plane. An incident EMW of frequency $f_{\rm EM}$ impinges the ROI section from the left hand-side along the $z$-axis and is transmitted to the opposite side's interface. The EM field spectrum is sampled in the following four points: $z_1$, $z_2$, $z_3$, and $z_4$. The different insets show the spectrum plot for each aforementioned point and each color refers to a sideband frequency.}
\label{fig:schematic_representation}
\end{figure*}

\subsection{Numerical setup and procedures}
\label{sec:simulation_conditions}
In this section, we describe and explain the configuration of the numerical experiments underlying our simulations, as well as the methods and the numerical implementation inherent to the FDTD technique.

The numerical experiment concerns about the interaction of a narrow-band EPW with a monochromatic GPW when both of them propagate along the same direction on the $z$-axis. In addition, we vary EPW properties, i.e., polarization, frequency, and amplitude, for finally studying how the GPW properties (polarization state amplitudes and frequency) affect them. In Fig.~\ref{fig:schematic_representation}, we display a scenario where the Region of Interest (ROI) extends along $0 \leq z \leq L$ (represented by the green-yellow-colored cuboid) and is filled with a material whose EM properties, i.e., $\overleftrightarrow \epsilon (z,t)$ and $\overleftrightarrow \mu(z,t)$, are described by Eq. \eqref{eq:subsection:Constitutive_relationships_for_non-covariant_Maxwell_equations_in_curved_spacetime_F}. The thickness of the ROI is set to $L = 100 \lambda_{\rm GW}$, being $\lambda_{\rm GW} = {c}/{f_{\rm GW}}$ the GPW wavelength. The remaining part of the space, i.e., regions $z<0$ and $z>L$, are defined by the EM properties of the vacuum, i.e., $\epsilon(z,t) = \epsilon_0$ and $\mu(z,t) = \mu _0$.

An EPW is generated from the left hand-side of the ROI, for instance, at the position $z_1$ shown in Fig.~\ref{fig:schematic_representation}. Regarding its spectral content, EPW must be narrow-band, i.e., with a bandwidth $\Delta f \leq f_{\rm GW}$ if $f_{\rm e} \gg f_{\rm GW} $, to properly appreciate the spectral features after the interaction. As it can be seen from Eqs. \eqref{eq:subsection:constitutive_relationship01} and \eqref{eq:subsection:constitutive_relationship02}, we can consider that the GPW modulates the EPW in amplitude a countless number of times, without changing its bandwith. Therefore, we use as source waveform a \emph{Gaussian pulse} modulated by a carrier frequency $f_{\rm e}$ expressed by
\begin{equation}
f(t,\theta,A,t_o,T) = A \sin(2\pi f_{\rm e} t+\Delta \theta) e^{-\left( \frac{t-t_o}{T}\right)^2},
\label{eq:waveform}
\end{equation}
where $A$ is the maximum EM field amplitude of the component, $\theta$ is a phase, $T$ is the Gaussian pulse width defined as $T= \frac{6 \sqrt{3}}{\pi f_{\rm GW}}$ to ensure that the 99\% of its energy is contained in the bandwidth $\Delta f = f_{\rm GW}/3$, and $t_{\rm o}= 3T$ is a delay used to avoid the introduction of numerical instabilities due to the source abrupt variations \cite{taflove2005computational}. Regarding the polarization properties of the source (or also known as the trace left by the electric field varying in time), it encompasses the following cases: linear on the $x$-axis (LX, horizontal) or $y$-axis (LY, vertical), linear and tilted about $45^\circ$ (L$45^\circ$) or $135^\circ$ (L$135^\circ$), Left Handheld Circular Polarization (LHCP), and Right Handheld Circular Polarization (RHCP), whose analytical expressions are all reported in Table \ref{table:incident_E_field}. 
\begin{table}[ht!]
\begin{ruledtabular}
\begin{tabular}{l l}
  Polarization & Incident electric field  (V/m)  \\ \hline
  LX    & $\vec{\mathcal{E}}_{\rm i}(z_s,t) = f(t,0,1,t_o,T)\vec{e}_{\rm x}$   \\ 
  LY    & $\vec{\mathcal{E}}_{\rm i}(z_s,t) = f(t,0,1,t_o,T)\vec{e}_{\rm y}$   \\ 
  L$45^\circ$    & $\vec{\mathcal{E}}_{\rm i}(z_s,t) = f(t,0,1,t_o,T)\vec{e}_{\rm x}+f(t,0,1,t_o,T)\vec{e}_{\rm y}$   \\ 
  L$135^\circ$    & $\vec{\mathcal{E}}_{\rm i}(z_s,t) = f(t,0,1,t_o,T)\vec{e}_{\rm x}-f(t,0,1,t_o,T)\vec{e}_{\rm y}$   \\ 
  LHCP    & $\vec{\mathcal{E}}_{\rm i}(z_s,t) = f(t,0,1,t_o,T)\vec{e}_{\rm x}+f(t,\frac{\pi}{2},1,t_o,T)\vec{e}_{\rm y}$ \\
  RHCP    & $\vec{\mathcal{E}}_{\rm i}(z_s,t) = f(t,\frac{\pi}{2},1,t_o,T)\vec{e}_{\rm x}+f(t,0,1,t_o,T)\vec{e}_{\rm y}$ \\
\end{tabular}
\end{ruledtabular}
\caption{Explicit mathematical expression of the electric field, $\vec{\mathcal{E}}_{\rm i}(z_s,t)$, of the incident EPW used as a source in our simulations. Here, $z_s$ is the source position, which is located on the left side of the ROI.}
\label{table:incident_E_field}
\end{table}

After a simulation is completed, we apply the Fourier transform to each component of the EM field already projected in the LIF to calculate the spectral properties of the EM field and its related polarization. The Fourier transform is computed at discrete frequencies given by $f_{\rm n} = f_{\rm e} + n f_{\rm GW}$, since the interference among sidebands in the frequency domain is expected to be negligible. Then, upon sampling the EM spectrum at the aforementioned frequencies, we obtain a global and complete vision of the interaction effects within the frequency domain. Figure \ref{fig:schematic_representation} shows several insets that report hypothetical changes in the EM field spectrum over the computational domain in the points $z_1$, $z_2$, $z_3$, and $z_4$. 

The study on the polarization properties of the EM field in the LIF is carried out via the \emph{Axial Ratio}, $AR$, defined as $AR=OA/OB$, where $OA$ and $OB$ are the major and minor axes (OB) of the trace left by the EPW on the $x-y$ plane over time, respectively.

The major and minor axes are given by \cite{balanis2005antenna}
\begin{subequations}
\begin{align}
OA &= \sqrt{\frac{\left|\mathcal{E}_x\right| ^2+\left|\mathcal{E}_y\right| ^2+G}{2}},\\
OB &= \sqrt{\frac{\left|\mathcal{E}_x \right| ^2+\left|\mathcal{E}_y \right| ^2-G}{2}},
\end{align}
\end{subequations}
where $G = \sqrt{\left| \mathcal{E}_x \right|^4+\left| \mathcal{E}_y\right|^4+2\left|E_x\right|^2 \left|E_y\right|^2 \cos\left( \Delta \phi\right)}$, $\mathcal{E}_x$ and $\mathcal{E}_y$ are the maximum magnitudes along the $x-$ and $y-$axes, and $\Delta \phi$ is the phase difference between $\mathcal{E}_x$ and $\mathcal{E}_y$. In Table \ref{table:axial_ratios}, we report the ranges of $AR$ in dB units for the three types of polarizations that an EMW can have, namely, circular, elliptical, and linear. Since we have a space-time varying material, $AR$ might change only along the $z$-axis, so we compute the average value
\begin{equation}
AR_{\rm Avg} = \sum _{k=1}^{N_{\rm ROI}} \frac{AR_{\rm k}}{N_{\rm ROI}},
\label{eq:ARaverage}
\end{equation}
where $AR_k=AR(z_k)$ is evaluated at the $k$-th cell inside the ROI and $N_{\rm ROI} = \lfloor L/\Delta z \rfloor$ denotes the number of cells contained inside the ROI with $\lfloor\cdot \rfloor$ representing the \emph{floor function}, see Sec.~\ref{sec:fdtdalgorithm}.
\begin{table}
\begin{ruledtabular}
\begin{tabular}{l l}
  Range of $AR$ & EMW polarization  \\ \hline
  $0 \leq \mathrm{AR} \leq 3$    & Circular  \\ 
  $3 \leq \mathrm{AR} \leq 25$    & Elliptical \\ 
  $25 \leq \mathrm{AR} < \infty$& Linear \\ 
\end{tabular}
\end{ruledtabular}
\captionsetup{justification=raggedright,singlelinecheck=off}
\caption{Typical ranges of $AR$, measured in dB, for different EMW polarization states.}
\label{table:axial_ratios}
\end{table} 

Finally, the grid used for the FDTD method possesses the following characteristics: cell length ${\Delta z = \frac{c}{10 f_{\rm max}}~=~6~{\rm mm}}$, where $f_{\rm max}= f_{\rm e}+40f_{\rm GW}$ is the maximum frequency expected to be computed with accuracy; time-step $\Delta T = {\rm CFLN}\frac{\Delta z}{ c} = 10~{\mu \rm{s}}$, where $\rm CFLN$ stands for the Courant-Friedrichs-Lewy number \cite{taflove2005computational} and is set to ${\rm CFLN} = 0.5$; GMIPML with polynomial profile degree $m=3$, normal theoretical reflection $R(0) = 10^{-7}$, and number of cells $N_{\rm pml} = 20$ \cite{zhao1998generalized}; total number of cells of the computational domain is set to $N_{\rm z} = 52940$, and the number of time steps is given by $N_{\rm ts} = 2N_{\rm z}+\lfloor5T/\Delta t \rfloor$. 

\section{Results}
\label{section:results_and_discussion}
In this section, we present the results of our FDTD numerical simulations. We compute the interaction of linear or circular polarized EPWs with a GPW endowed with only one of the Linear Polarization states between $h_{\oplus}$ (see Sec. \ref{sec:GPW_hplus}) and $h_{\otimes}$ (see Sec. \ref{sec:GPW_hcross}). 

\subsection{GPWs with $h_{\oplus}$ polarization state}
\label{sec:GPW_hplus}
\begin{figure*}[ht!]
\begin{subfigure}{0.30\linewidth}
\centering
\includegraphics[scale = 0.35]{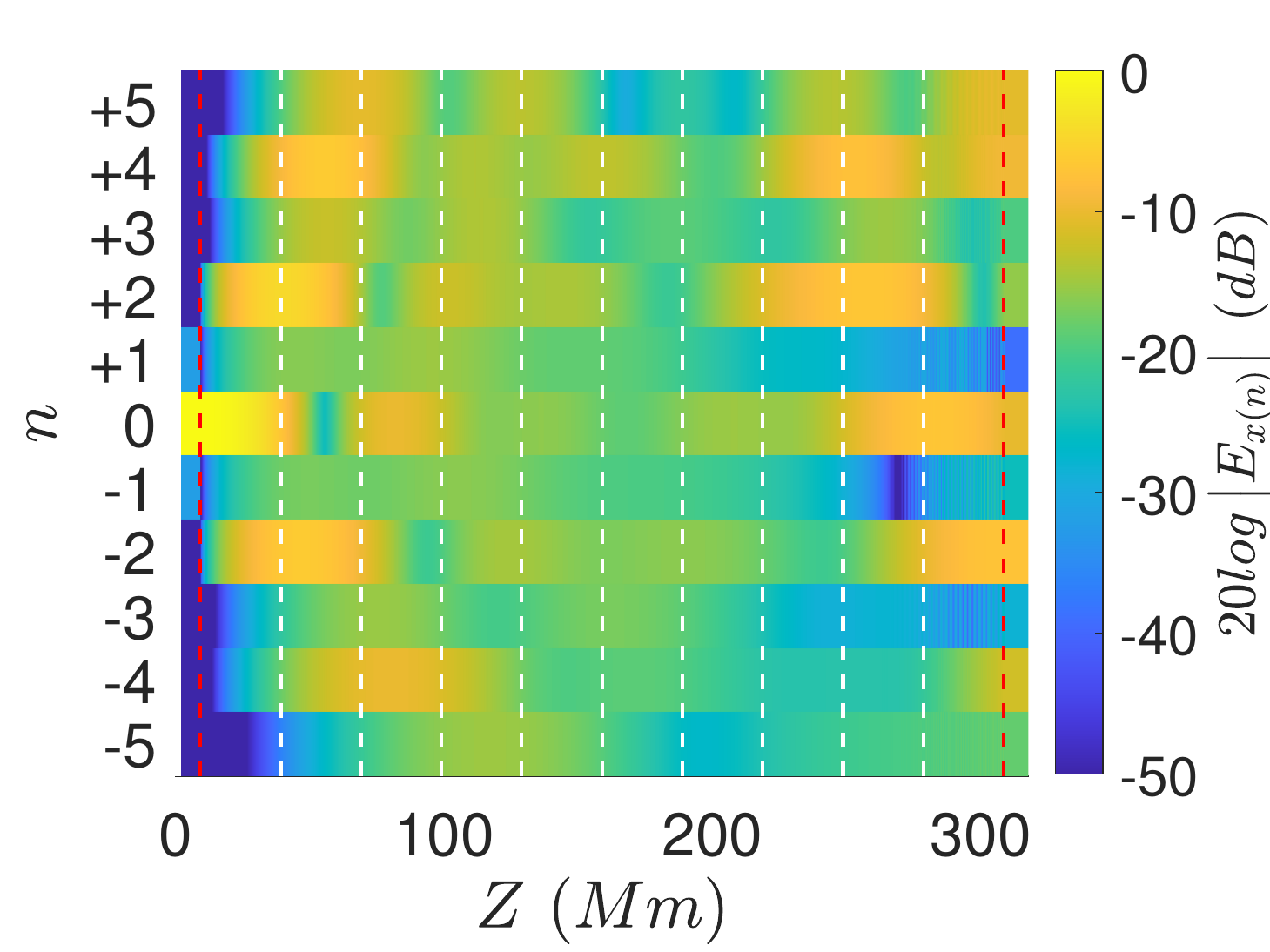}
\caption{$h_{\oplus} = 0.1$, $f_{\rm GW} = 100~Hz$}
\label{figure:results_EMX_h+_a}
\end{subfigure}\hfill
\begin{subfigure}{0.30\linewidth}
\centering
\includegraphics[scale = 0.35]{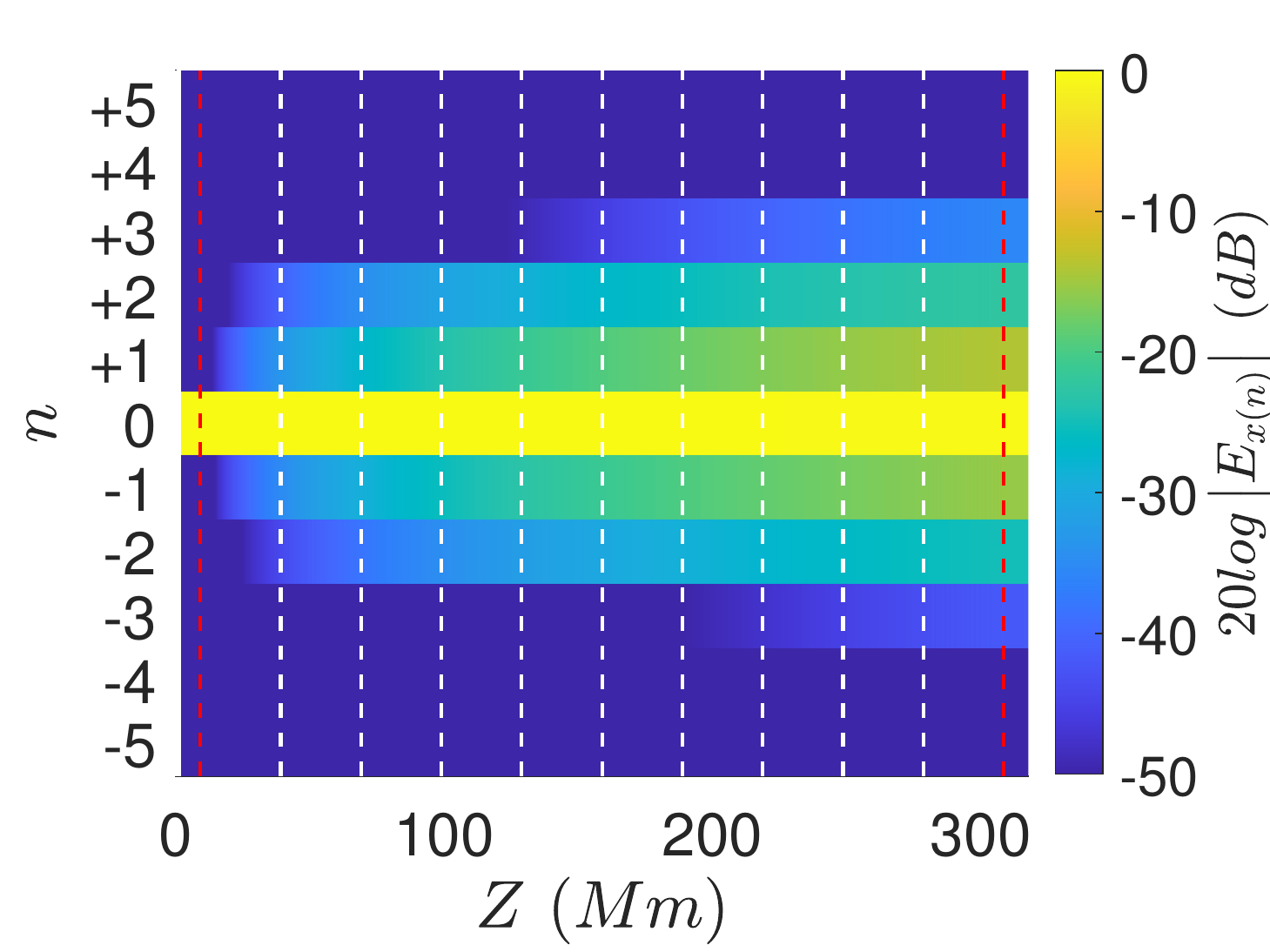}
\caption{$h_{\oplus} = 0.01$, $f_{\rm GW} = 100~Hz$}
\label{figure:results_EMX_h+_b}
\end{subfigure}\hfill
\begin{subfigure}{0.30\linewidth}
\centering
\includegraphics[scale = 0.35]{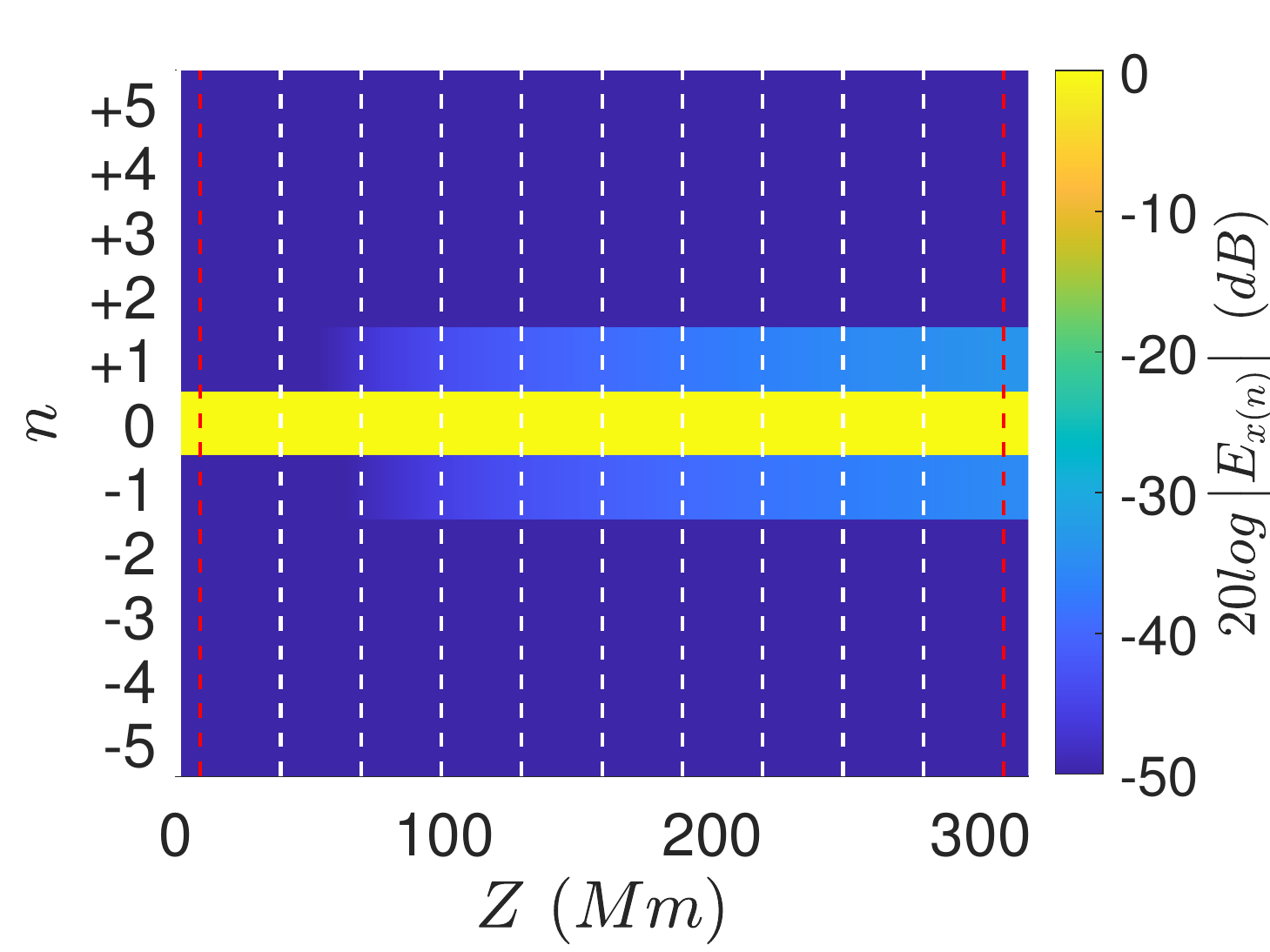}
\caption{$h_{\oplus} = 0.001$, $f_{\rm GW} = 100~Hz$}
\label{figure:results_EMX_h+_c}
\end{subfigure}\hfill
\begin{subfigure}{0.30\linewidth}
\centering
\includegraphics[scale = 0.35]{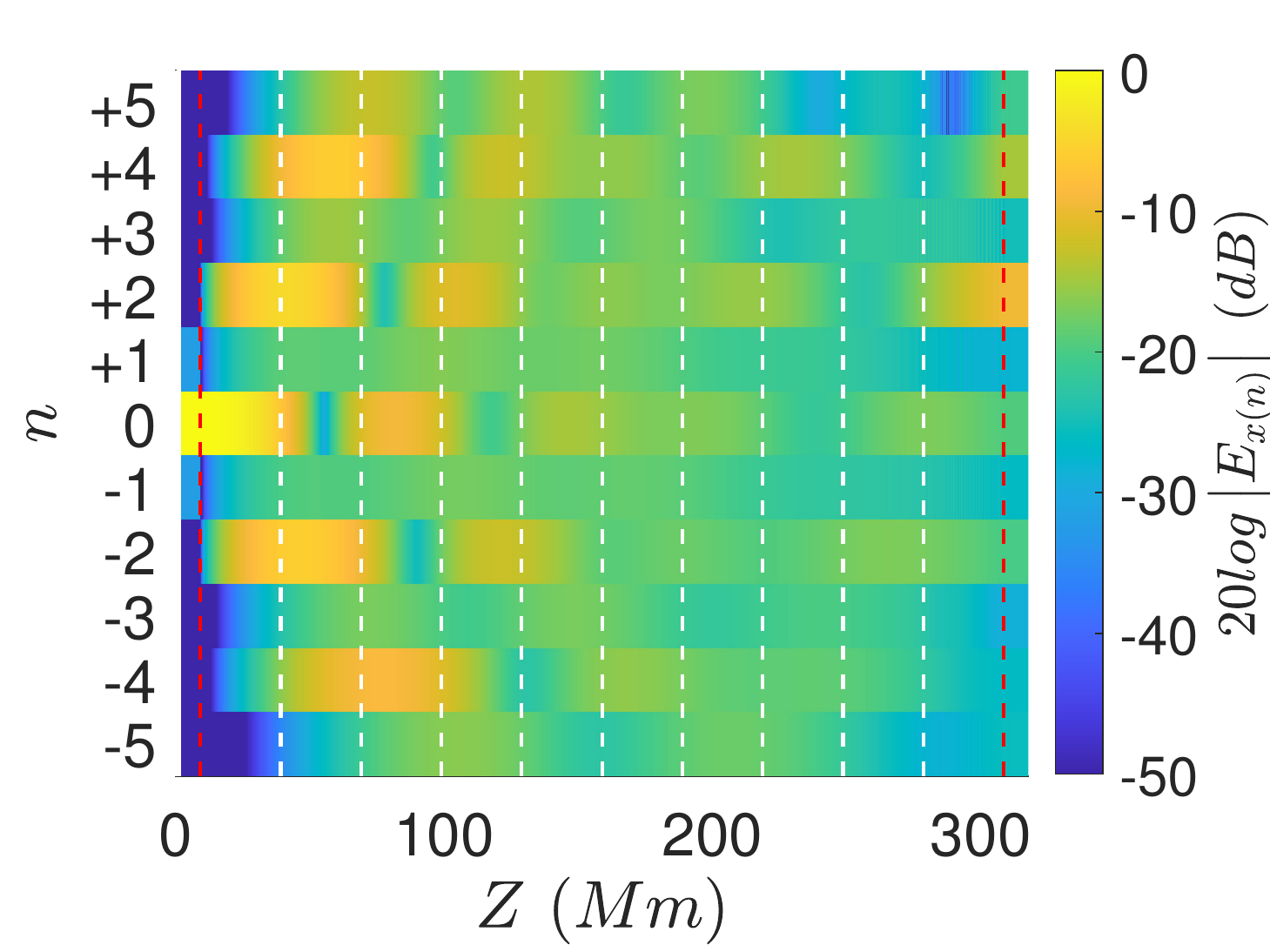}
\caption{$h_{\oplus} = 0.1$, $f_{\rm GW} = 75~Hz$}
\label{figure:results_EMX_fgw_2a}
\end{subfigure}\hfill
\begin{subfigure}{0.30\linewidth}
\centering
\includegraphics[scale = 0.35]{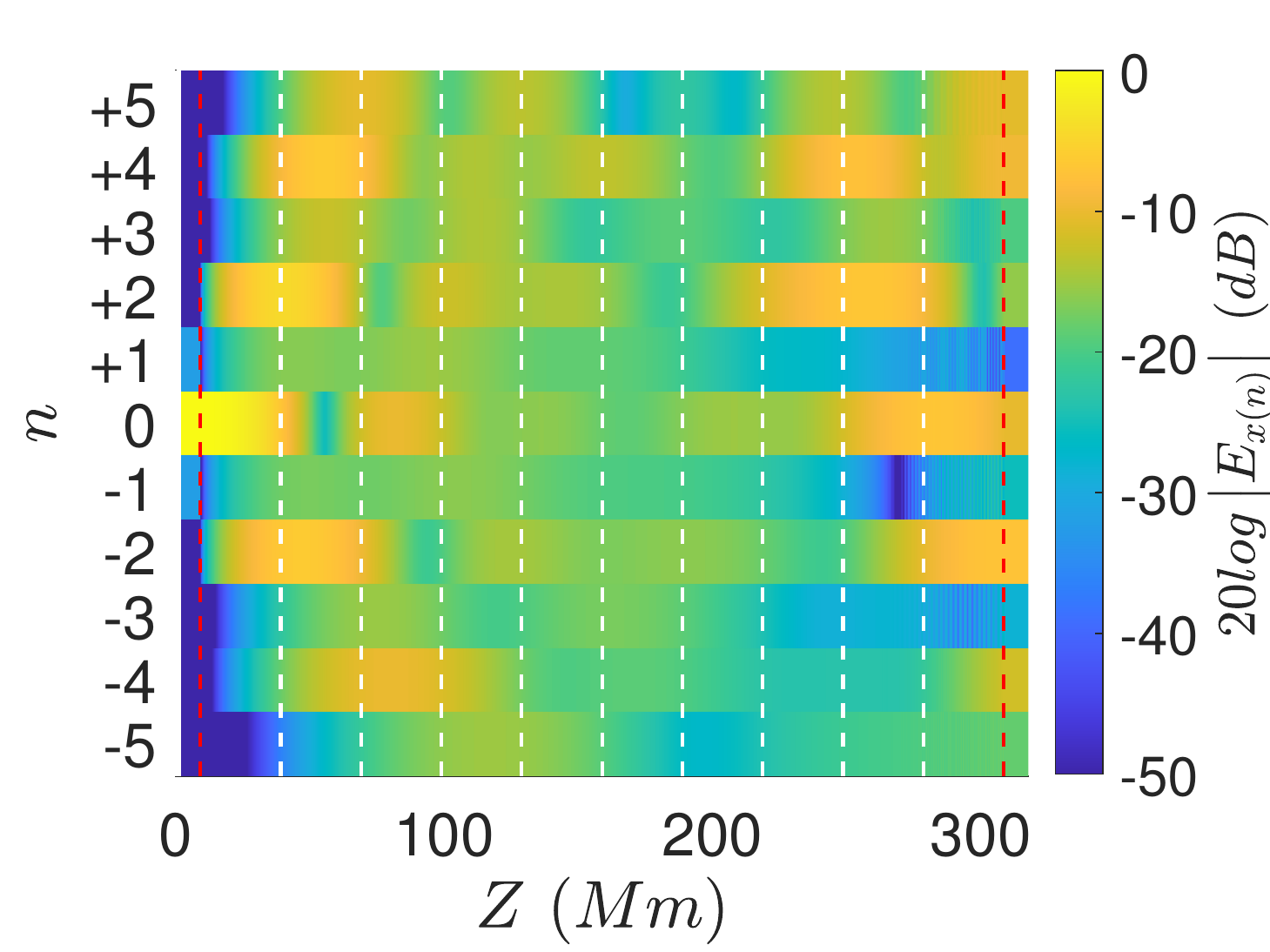}
\caption{$h_{\oplus} = 0.1$, $f_{\rm GW} = 100~Hz$}
\label{figure:results_EMX_fgw_1a}
\end{subfigure}\hfill
\begin{subfigure}{0.30\linewidth}
\centering
\includegraphics[scale = 0.35]{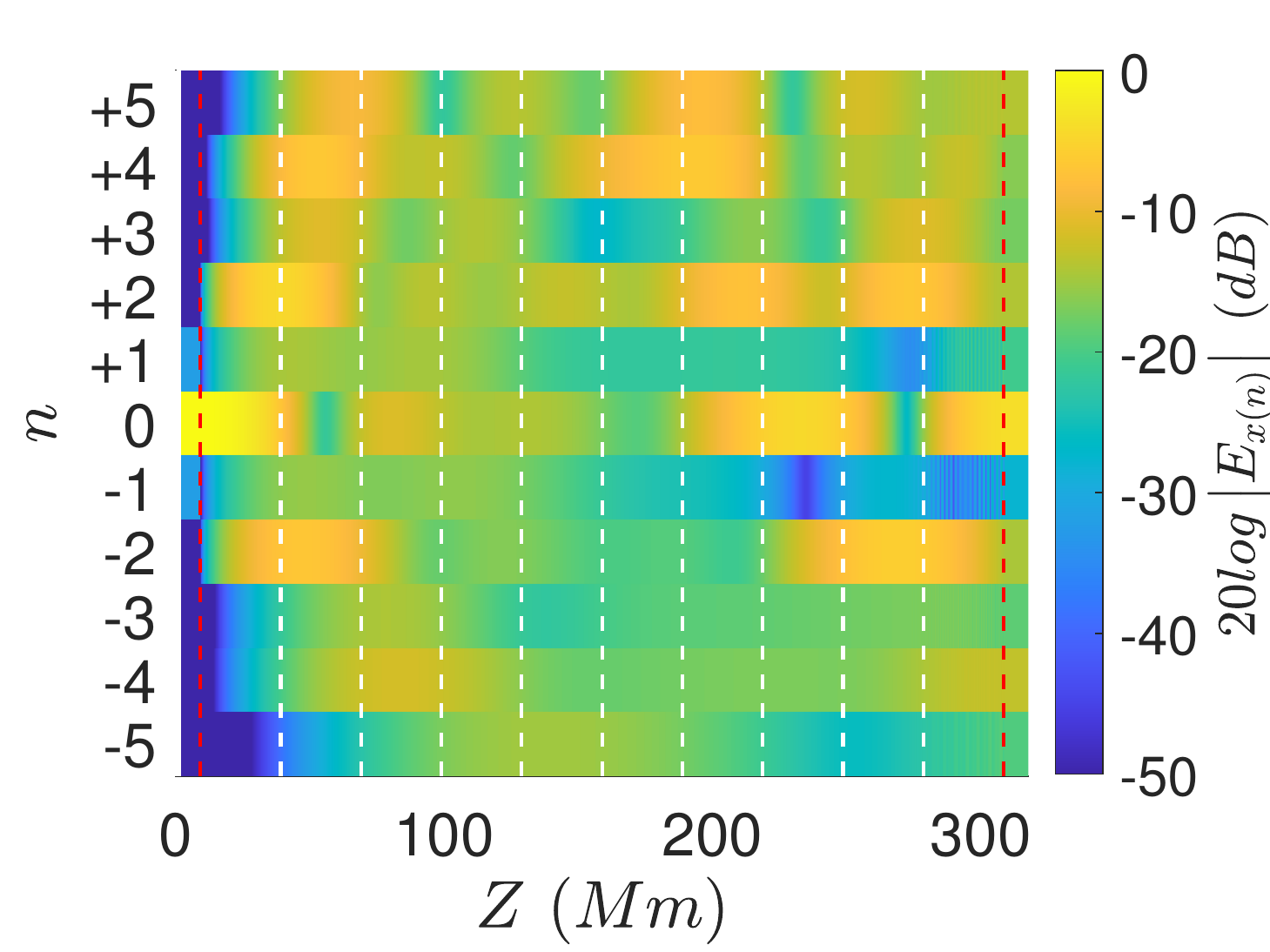}
\caption{$h_{\oplus} = 0.1$, $f_{\rm GW} = 125~Hz$}
\label{figure:results_EMX_fgw_2b}
\end{subfigure}\hfill
\caption{Spatial distribution of the $x$ component of the electric field for 10 sidebands. The dashed vertical lines delimits the results calculated in the LIF. The lateral color map is expressed in dB units and normalized with respect to the maximum magnitude attained at the $n = 0$ polarization state.}
\label{figure:results_EM_set01}
\end{figure*}

We first consider a GPW with $h_{\oplus} \neq 0$ and $h_{\otimes}=0$, which interacts an incident EPW with polarization LX (see Table. \ref{table:incident_E_field}). The results of these numerical experiments are displayed in Fig.~\ref{figure:results_EM_set01}, where each plot shows the evolution of 11 spectral-components at frequencies $f_n = f_e + n f_{\rm GW}$ for different values of GPW parameters $h_{\oplus}$ and $f_{\rm GW}$, where $n = 0, \pm 1, \pm 2 ,\ldots,\pm 5$ (vertical axes) related to the magnitude of the $x$-component of the electric field (expressed in dB units in the lateral color map) as seen by an observer in rest at each point of the $z$-axis (horizontal axes measured in megameter units, i.e., $1~\rm{Mm} = 1000~\rm{km}$). Here, the ROI (yellow-green colored block shown in Fig.~\ref{fig:schematic_representation}) is represented in each plot of Fig.~\ref{figure:results_EM_set01} as the region delimited between the dashed red lines. In this case, no $y$ component is excited, since the EM properties defined in Eq. \eqref{eq:subsection:Constitutive_relationships_for_non-covariant_Maxwell_equations_in_curved_spacetime_F} have no off-diagonal terms under the current simulation conditions. Therefore, there is no coupling between either $x$- or $y$-components of the electric field neither in the standard frame nor in the LIF.

The results shown in the first row of Fig.~\ref{figure:results_EM_set01} are obtained by varying the GPW amplitude as follows $h_{\oplus}=0.1,~0.01,$ and $0.001$ (see panels (a), (b), and (c), respectively), while GW frequency is held constant at $f_{\rm GW} =100~\text{Hz}$. Therefore, when the EPW passes through the ROI, it interacts with the space-time-varying region endowed with certain EM properties (i.e., the analogue EM-model of a GPW). In addition, its spectrum is progressively modulated at each point of the $z$-axis, exciting then several sidebands at frequencies $f_n = f_{\rm e} \pm n f_{\rm GW}$. These progressive up- and down-conversion processes can be interpreted as a multiple scattering of photons by gravitons in the GR realm \cite{kopvillem1973detection}; whereas in the telecommunication framework, it shares deep similarities with the cascade Brillouin scattering effect \cite{wolff2017cascaded}, generated in photonic crystal fibers or in non-reciprocal microwave devices based on isotropic materials with space-time-varying EM properties \cite{taravati2017nonreciprocal,taravati2018giant}. This situation is schematically illustrated in Fig.~\ref{fig:schematic_representation}, where the insets show samples of the electric field spectrum at different positions inside the grid. This is the the most remarkable difference among the proposed analogies, namely the number of spectral components present in each one. Despite we only display 11 spectral components, the breadth of the effect is wider since if the EPW propagated in an unlimited GPW space-time background, i.e, $L \to \infty$, an infinite number of sidebands would be produced, but their amplitudes would be expected to be extremely small, but not null. 

Panels (a), (b), and (c) of Fig.~\ref{figure:results_EM_set01} show how the influence of the GPW amplitude is responsible for the increment or reduction of the spatial rate at which the energy is exchanged between sidebands. The larger the GPW amplitude, the shorter the interaction length will be, such that even higher-order sidebands appear. In addition, from our simulations we deduce that even-order sidebands have higher amplitudes than odd-order ones. 

Instead, panels (d), (e), and (f) of Fig. \ref{figure:results_EM_set01} show the results when the GPW frequency is varied as $f_{\rm GW} = 75,100,$ and $125$ Hz, while its amplitude is kept constant to $h_{\oplus} =0.1$. Again, as it occurred when the GPW amplitude was varied, several sidebands appear, but similarly, the GW frequency affects the spatial rate with which the sidebands are excited. In this case, the larger the GPW frequency, the shorter the necessary interaction length will be for energy transfer among sidebands. 

We now discuss the EM polarization effects not shown in Fig. \ref{figure:results_EM_set01}. We first note that the polarization of each sideband keeps linear when the incident EPW is LX or LY polarized. Instead, when the incident EPW has either LHCP or RHCP polarization, then the EM field components for the even sidebands will keep the same polarization of the incident wave, whereas the odd sidebands will be either RHCP or LHCP polarized, respectively. Finally, when the incident EPW is a superposition of LX and LY, i.e., L$45^\circ$ or L$135^\circ$, the polarization of each sideband is linear, but again, even sidebands keep the polarization of the incident EPW, while odd sidebands are out of phase by $\pi$. 

\subsection{GPW with $h_{\otimes}$ polarization state}
\label{sec:GPW_hcross}
In this section, we concentrate on the effects induced by a GPW with $h_{\otimes} = 0$ and $h_{\otimes} \neq 0$. We found that the effect of both GPW amplitude, $h_{\otimes}$, and frequency, $f_{\rm GW}$, are same as those explained in the previous section. In other words, a set of sidebands are generated due to the modulation effects produced by the EM-analogue model of the GPW and the production-rate is influenced by the GPW amplitude and frequency. 

\begin{figure*}[ht!]
\begin{subfigure}{0.30\linewidth}
\centering
\includegraphics[scale = 0.35]{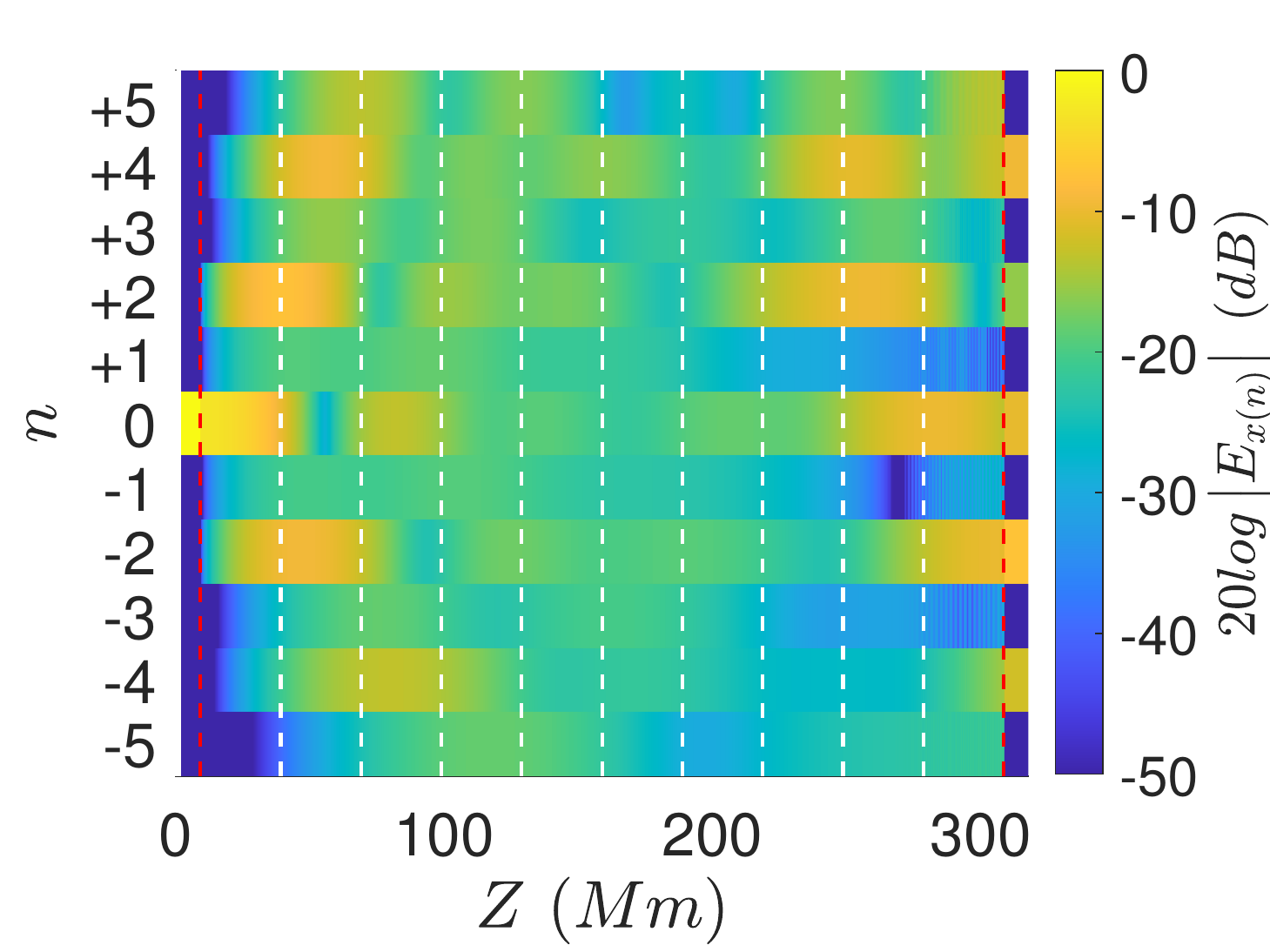}
\caption{POL: LPX}
\label{figure:results_EMX_hx_a}
\end{subfigure}\hfill
\begin{subfigure}{0.30\linewidth}
\centering
\includegraphics[scale = 0.35]{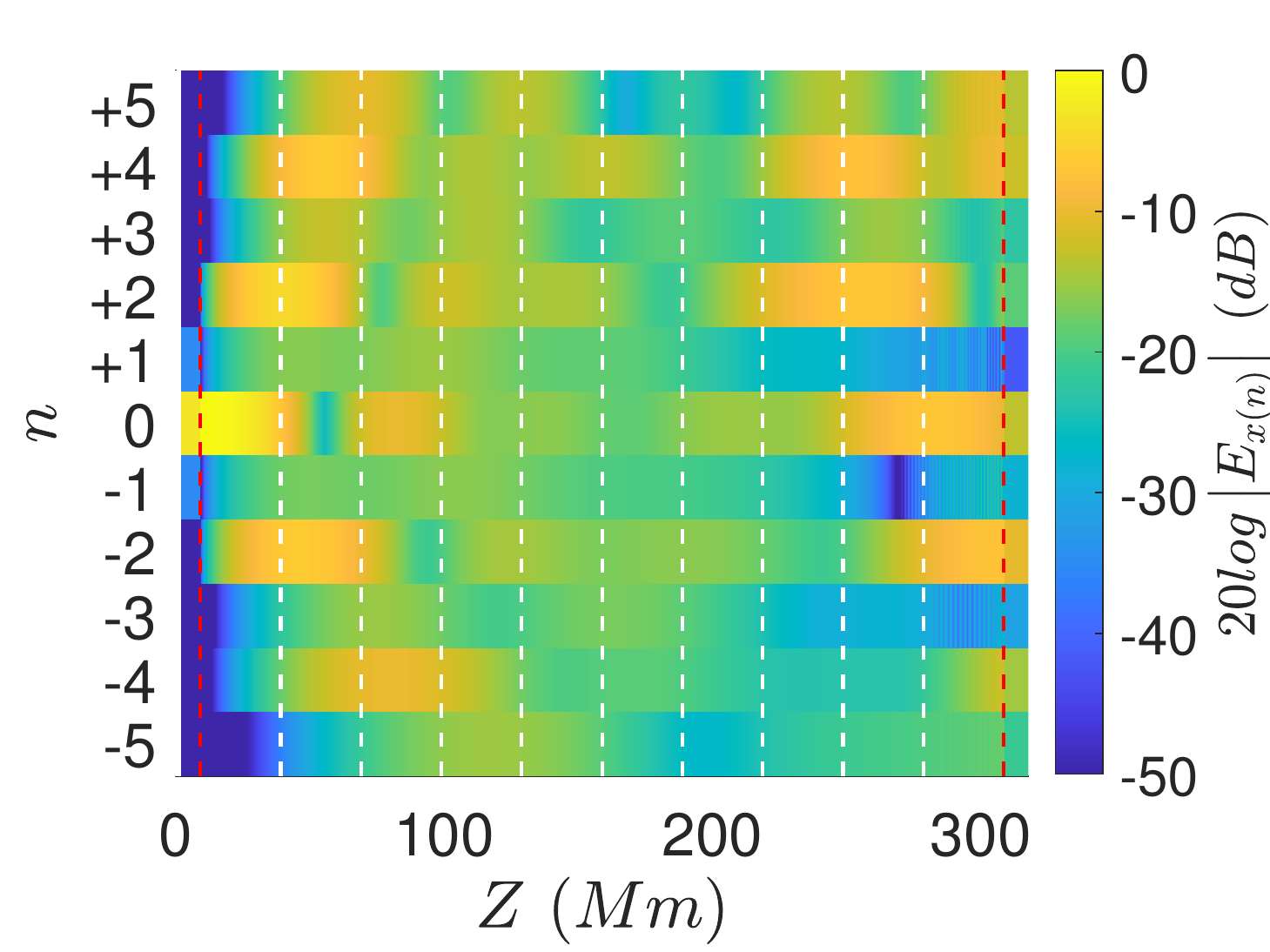}
\caption{POL: LP-45$^\circ$}
\label{figure:results_EMX_hx_b}
\end{subfigure}\hfill
\begin{subfigure}{0.30\linewidth}
\centering
\includegraphics[scale = 0.35]{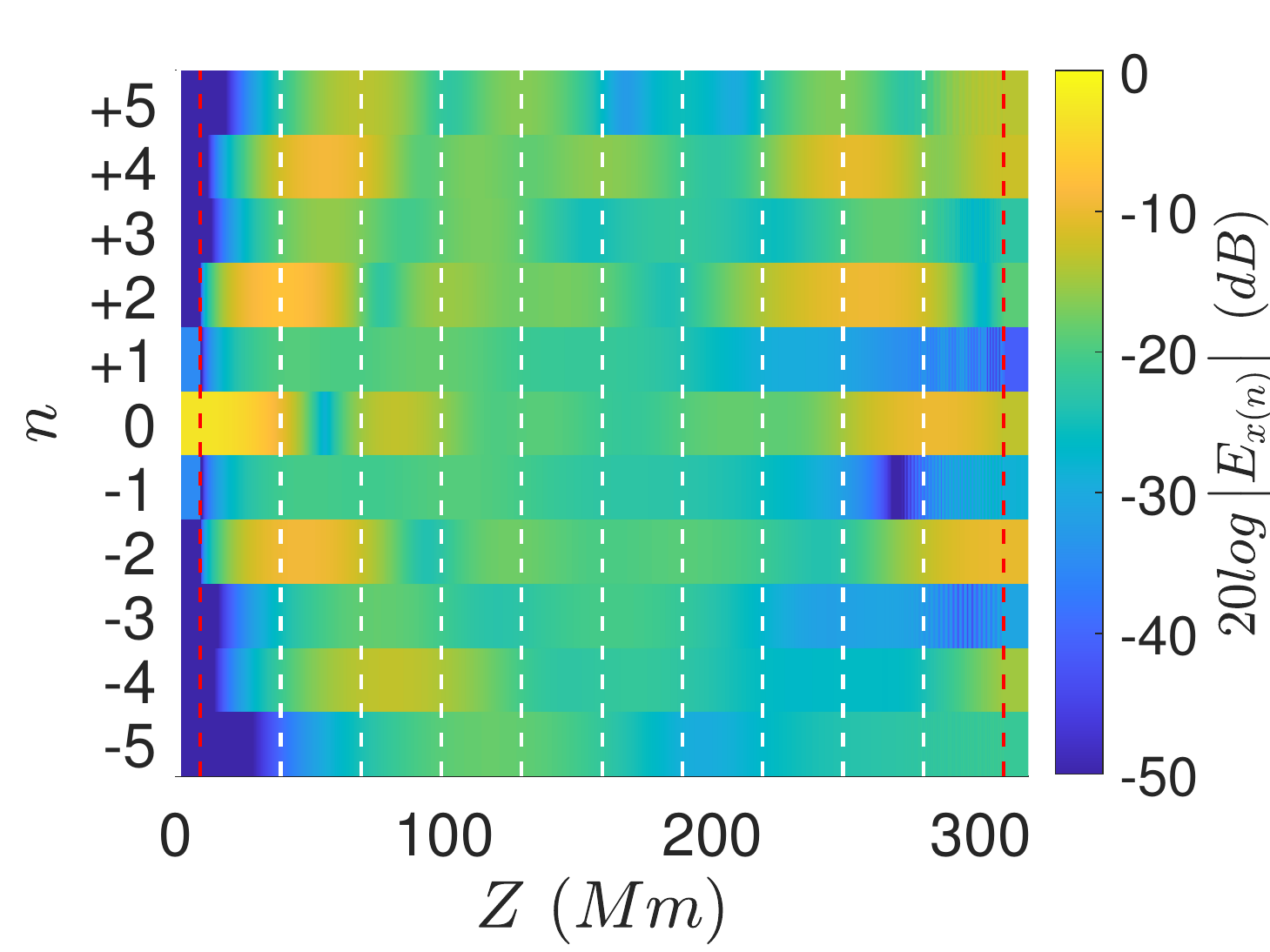}
\caption{POL: LHCP}
\label{figure:results_EMX_hx_c}
\end{subfigure}\hfill
\begin{subfigure}{0.30\linewidth}
\centering
\includegraphics[scale = 0.35]{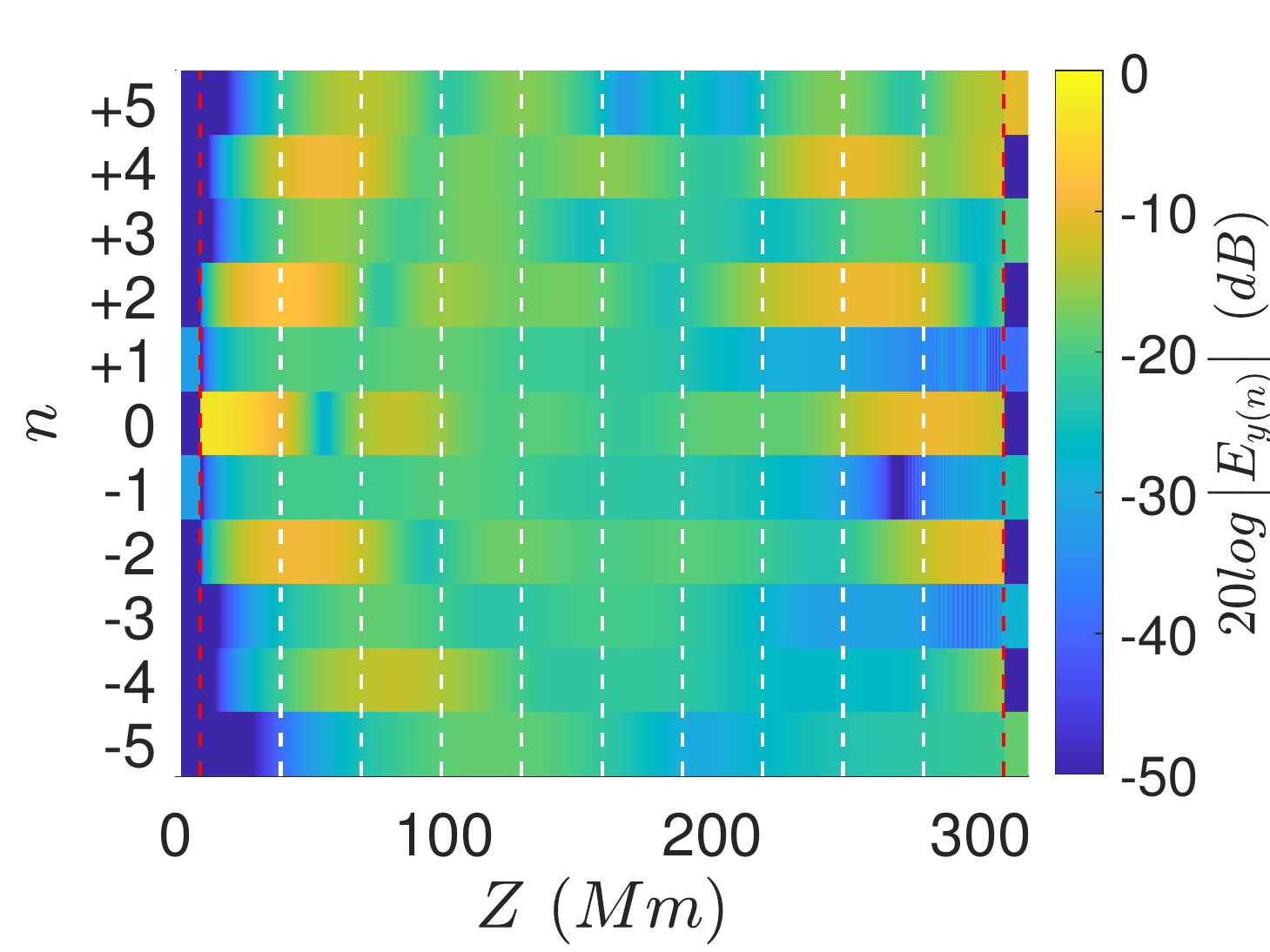}
\caption{POL: LPX}
\label{figure:results_EMY_hx_a}
\end{subfigure}\hfill
\begin{subfigure}{0.30\linewidth}
\centering
\includegraphics[scale = 0.35]{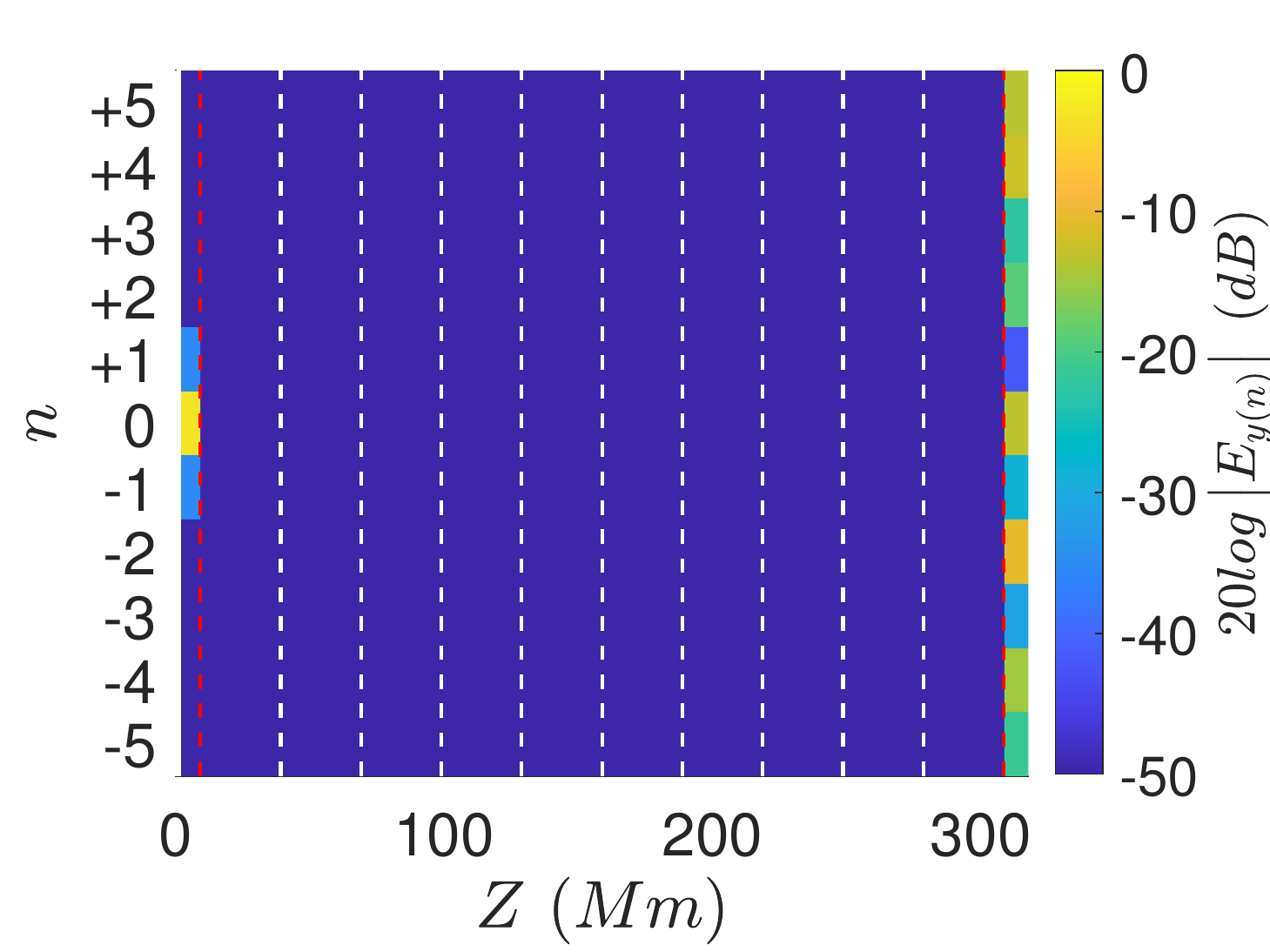}
\caption{POL: LP-45$^\circ$}
\label{figure:results_EMY_hx_b}
\end{subfigure}\hfill
\begin{subfigure}{0.30\linewidth}
\centering
\includegraphics[scale = 0.35]{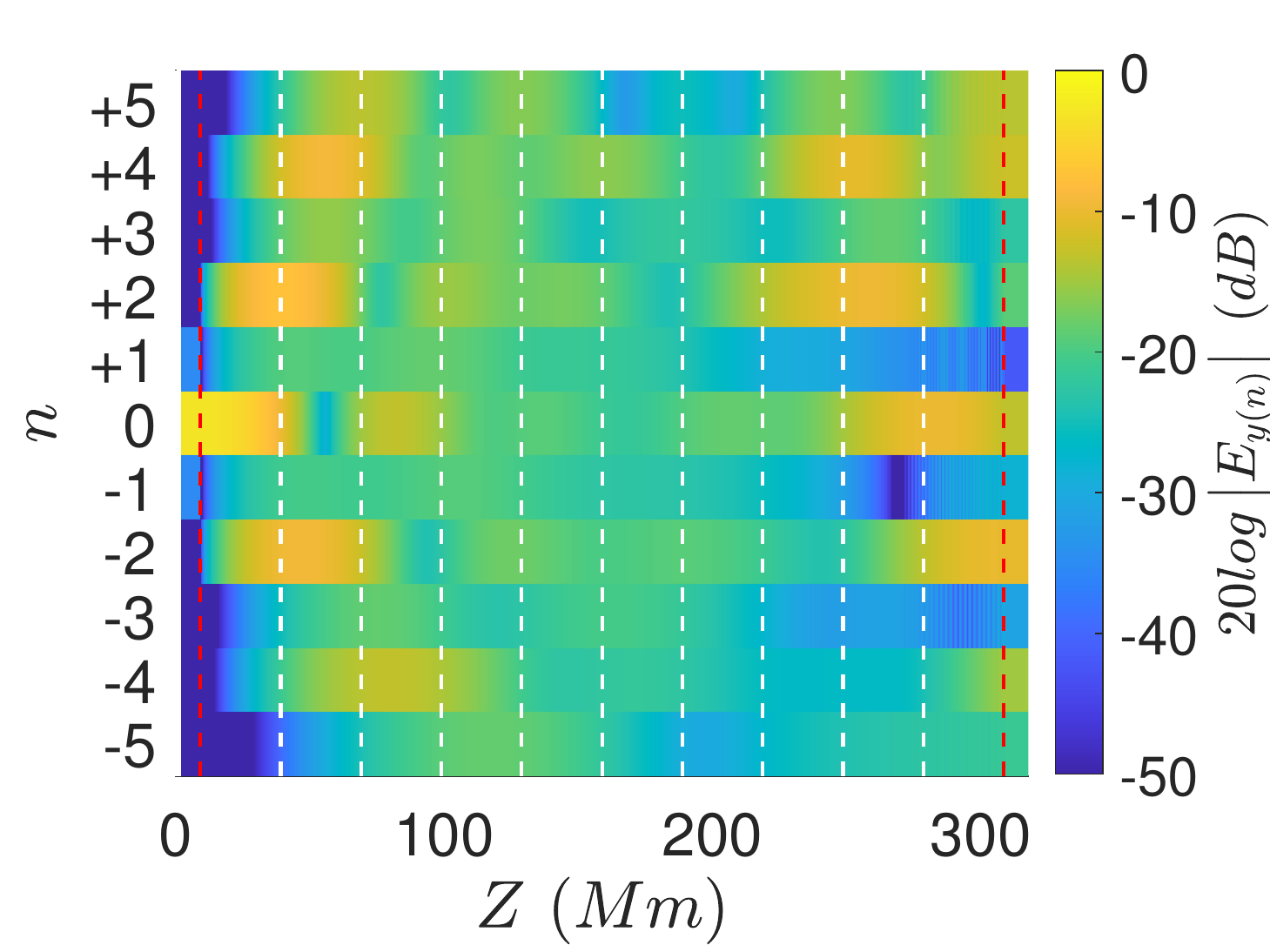}
\caption{POL: LHCP}
\label{figure:results_EMY_hx_c}
\end{subfigure}\hfill
\caption{Spatial distribution of the $x$- and $y$-components of the electric field for 10 sidebands and for several polarization states of the incoming EMW. We employ the same notations of Fig. \ref{figure:results_EM_set01}. GPW characteristics are $h_{\otimes}=0.1$ and $f_{\rm GW} =100~Hz$. The symbol ``POL'' stays for polarization.}
\label{figure:results_EM_hx_different_cases}
\end{figure*}

Therefore, we focus on the effect that the GPW has on the incident EMW polarizations LX, L$45^\circ$, and LHCP. Since the EM-analogue model of gravity (cf. Eq. \eqref{eq:subsection:Constitutive_relationships_for_non-covariant_Maxwell_equations_in_curved_spacetime_F}) presents off-diagonal components in both the dielectric permittivity and magnetic permeability tensors, then there is a coupling between the EM field components ${\mathcal{E}_x}$ and $\mathcal{E}_y$ both in the standard and LIF frames. In this tranche of simulations, the GPW amplitude and frequency are set to $h_{\otimes} =0.1$ and $f_{\rm GW} = 100$ Hz, respectively. The first and second rows of Fig. \ref{figure:results_EM_hx_different_cases} again show the spatial evolution of the magnitude of 11 sidebands for the $x$- and $y$-components of the electric field, respectively. Each column corresponds to a polarization state of the incident EPW: LX (panels (a) and (d)), L$45^{\circ}$ (panels (b) and (e)), and LHCP (panels (c) and (f)).

When the polarization of the incident wave is LX, we have that the observers in the LIF experience an electric field with both $\mathcal{E}_x$ and $\mathcal{E}_y$ components, but the polarization remains still linear. In addition, we discover that there is a phase difference of $\pi$ between the even and odd sidebands. On the other hand, for an incident EPW with polarization L$45^\circ$, there is no $y$-component excited in the LIF, since given the symmetry of the problem, the electric field in the standard coordinate system lays along the direction of the tetrad $\tau^\mu _{(1)}$ (settled by us along the $x$-axis in the LIF). We can say that there is a total destructive interference effect for the $y$-component of the EM field. 

If the incoming EPW has polarization L$135^\circ$ (not shown in Fig.~\ref{figure:results_EM_hx_different_cases}), then in the LIF only the $y$-component will be observed. However, this time the reason behind this effect is due to the fact that the polarization of the incident EPW in the standard coordinate system lays on the direction of the tetrad $\tau^\mu _{(2)}$, which corresponds in our conventions to the $y$-axis in the LIF.

Finally, we consider the case where the incident EPW has LHCP polarization. Similar to the situation described in the previous section, where there is an alternation of the polarizations among even and odd sidebands with respect to the fundamental band ($n = 0$), now even sidebands keep the polarization of the incident wave, whereas odd sidebands have RHCP polarization. If the polarization of the incident wave is RHCP (not shown in  Fig.~\ref{figure:results_EM_hx_different_cases}), the perpendicular polarization (cross-polar component) manifests. It is worth mentioning that from our simulations emerges that the amplitude of the odd sidebands is in general lower than the even ones for all the studied cases.

\section{Conclusions}
\label{section:conclusion}
The interaction between a monochromatic GPW of frequency $f_{\rm GW}$ and a narrow-band EPW with carrier frequency  $f_{\rm e}$, respectively, when propagating in the same direction, generates infinite sidebands in the EM spectrum, having frequencies $f_{\rm n} = f_{\rm e}+n f_{\rm GW}$, where $n$ is an integer number corresponding to the order of the sideband. The phenomenon is similar to the cascade Brillouin effect between sidebands \cite{wolff2017cascaded}. This may be interpreted as multiple scattering of a photon by gravitons. When the GPW and EPW are propagating in opposite direction, no interaction is produced. The effect is present for both $LP-h_{\oplus}$ and $LP-h_{\otimes}$ polarization states of the GPW. Amplitude and frequency of the GPW define the amplitude and spatial rate of sideband excitation.

Studying the GW properties by means of the polarization of the induced sidebands is feasible, but can be also quite subtle. For example, when an EM with $LP-45^\circ$ polarization is incoming, no $y$-component will be generated at the sidebands. This effect could be understood as having $LP-h_{\oplus}$ GPW, which is not the case. This problem might be solved when the GPW and the EPW propagate neither in the same direction nor in opposite ones, but they form a specific angle. This case deserves a detailed study in a separate paper. 

Finally, it is worth mentioning that either $h_{\oplus}$ or $h_{\otimes}$ values for the GWs proposed here are not realistic, since they are very large compared with the physical ones (e.g., see Refs. \cite{abbott2017gw170608,abbott2019gwtc,abbott2016gw151226,ligo2017basic}, for details). We have used high numerical numbers, only for testing the cascade Brillouin effect. However, if we refer to the astrophysical event GW150914, constituted by two merging black holes \cite{ligo2017basic,postnov2014evolution,huwyler2014testing}, this generates a GW with $h_{\oplus} = 10^{-6}$ and $f_{\rm GW} = 100~\rm Hz$ at a distance of $0.157~\rm au$. If we consider $f_{\rm e}=1~\rm kHz$, then two sidebands will be generated with amplitudes of around $-100~\rm dB$ with respect to the $n=0$ band. 

{Since the breadth of the effect under study depends on the interaction-length (distance along which both GW and EPW interact), simulations involving lower GWs amplitudes ($h_{\oplus}, h_{\oplus} \leq 10^{-6}$) are constrained by of computational resources. Therefore, if the interaction length is not long enough, then sidebands will be masked by the spectrum of the source. However, we remark that the amplitude of sidebands are not null. In addition, although these levels are extremely weak for being measured with the present observational technology, future higher sensitivity instruments might be able to measure them.}

The effect we have studied in this paper represents an alternative and original mechanism to indirectly detect GW emissions by means of nearby luminous stars or accreting sources. Indeed, the variations of their EM X-ray radiation (being astrophysically related to changes in luminosity or flux), albeit are very small, can be used as natural interferometer instruments. Indeed, their EM emission can be analysed by exploiting standard methods in X-ray astronomy, together with the scheme presented in this work. Depending where such brilliant astrophysical objects are located with respect to the GW source, it is related to the amplitude of variations present in their EM spectrum. However, generally to appreciate these changes we need higher observational sensitivity, that may be reached with the future upgrades. 

\begin{acknowledgments}
The authors would like to thank MARTINLARA Project  P2018/NMT-4333. V. D. F. acknowledges Istituto Nazionale di Fisica Nucleare (INFN), Sezione di Napoli, \textit{iniziative specifiche} TEONGRAV, and Gruppo Nazionale di Fisica Matematica of Istituto Nazionale di Alta Matematica.
\end{acknowledgments}

\section*{Data Availability Statement}
The data that support the findings of this study are available from the corresponding author upon reasonable request.

\bibliography{references}

\end{document}